\documentclass[amsmath,amssymb,aps,floatfix,twocolumn]{revtex4-1}

\usepackage{graphicx}
\usepackage{bm}
\usepackage{nicefrac}
\usepackage{xcolor}

\DeclareMathOperator{\cov}{\text{cov}}

\let\ddt=\relax
\DeclareMathOperator{\ddt}{\frac{\text{d}}{\text{dt}} \!}

\begin{document}


\title{Inferring stochastic regulatory networks from perturbations of the non-equilibrium steady state}

\author{Niklas Bonacker and Johannes Berg}
\email{nbonacke@uni-koeln.de and bergj@uni-koeln.de}
\affiliation{Institute for Biological Physics, University of Cologne,
  Z\"ulpicher Stra{\ss}e 77a, 50937 Cologne, Germany}
\date{\today}

\begin{abstract}
Regulatory networks describe the interactions between molecular or cellular regulators, like 
transcription factors and genes in gene regulatory networks, kinases and their receptors in signalling networks, or neurons in neural networks. A long-standing aim of quantitative biology is to reconstruct such networks on the basis of large-scale data. Our aim is to leverage fluctuations around the non-equilibrium steady state for network inference. To this end, we use a stochastic model of gene regulation or neural dynamics and solve it approximately within a Gaussian mean-field theory. We develop a likelihood estimate based on this stochastic theory to infer regulatory interactions from perturbation data on the network nodes. We apply this approach to artificial perturbation data as well as to phospho-proteomic data from cell-line experiments and compare our results to inference schemes restricted to mean activities in the steady state. 
\end{abstract}

\maketitle


\section{Introduction}

All forms of life have developed remarkable mechanisms to process information.
For instance, \emph{neural networks} process sensory information and achieve motor control, among many other feats. \emph{Gene regulatory networks} integrate both external and internal signals into patterns of gene expression, computing for instance responses to external stresses. They consist of molecular regulators such as transcription factors (proteins that can bind to specific sites on DNA and affect the expression of genes, including those that encode transcription factors themselves). Gene regulatory networks also orchestrate the spatio-temporal patterning that is central to embryo development.
\emph{Signalling networks} consist of protein kinases that phosphorylate and regulate the activity of other enzymes, channels, and molecular transporters. Signalling networks typically form small pathways of tens of kinases and perform information processing for development, tissue repair, and cell death. Gene regulation and signalling are often tightly integrated through the phosphorylation, and thus activation, of transcription factors.

These different types of regulatory networks --- neural networks, gene regulatory networks and signalling networks --- share three essential characteristics: they are based on (i) complex topological structures described my a matrix of interactions, their components are (ii) intrinsically stochastic, and they (iii) operate out of equilibrium.

(i) In neural networks the nodes correspond to neurons, in the case of gene regulatory networks they correspond to genes, and in signalling networks to proteins. Each node is characterized by an activity, which would be the firing state in the case of a neuron, or the concentration of different gene products and different phosphorylation states for gene regulatory networks and signalling networks. These activities change over time, depending on the activities of other nodes and on external signals. The effect of the activity of one node $i$ on another $j$ is quantified by a regulatory interaction between nodes.

(ii) Neural activity is stochastic because the transmission of a signal across a synapse is effected by the stochastic release of a neurotransmitter. Analogously, both the dynamics of gene expression and signalling are stochastic due to the small number of molecules involved in ligand binding~\cite{Raj2008}.

(iii) Biological information processing operates out of equilibrium because the regulatory networks describing the connections between neurons, genes, or kinases are generally asymmetric: Neuron $i$ giving input to neuron $j$ does not imply neuron $j$ gives input to neuron $i$ and vice versa for gene regulation and signalling. As a result, the steady state of the network is a non-equilibrium steady state; there exists no energy or free energy function which is minimized by the dynamics of the regulatory network. The steady state of the activities does not obey detailed balance and is not described by a Boltzmann distribution.

In this paper we ask how well regulatory networks can be inferred on the basis of data on the activity of nodes. The motivation for \emph{network inference} is twofold: Determining regulatory interactions experimentally remains an arduous task; except for well-studied pathways and individual transcription factors, regulatory networks are poorly characterized. On the other hand, it is comparatively easy to collect large amounts of data on the activity of nodes (through high-throughput methods such as microarray techniques~\cite{Hoheisel2006} or RNA-seq~\cite{Wang2009,Chu2018,Alvis2018} for gene expression and multi-electrode techniques~\cite{kodandaramaiah2018multi} or electrode arrays \cite{obien2015revealing} for neurons). The inference of regulatory networks given the activity of nodes is an inverse statistical problem~\cite{Nguyen2017} with a large body of work in different disciplines~\cite{Nee2004,Schneidman2006a,Margolin2006,Meyer2007,zhu2008integrating,Hecker2009,Cocco2009a,Hill2012,Li2015}.

A key feature of regulatory networks making their inference a challenging problem is the network asymmetry. This can be illustrated with a simple argument: The number of potential interactions between $N$ nodes is $N(N-1)$ for a general asymmetric interaction matrix without self-interactions. However, the matrix of covariances is symmetric, and is described by only $N(N+1)/2$ parameters. To infer the interactions between nodes from covariances alone is thus not possible. This is a particularly egregious example of the \emph{curse of dimensionality} well known in inference problems, in the sense that even if observables like the covariances can be determined accurately from the data this is insufficient for network inference. In principle, one can use cumulants of higher order (for instance triplets of nodes), but it turns out that, since higher cumulants are often numerically small, they require very large datasets to be determined with sufficient accuracy~\cite{dettmer2016network}.

To address this problem, we build on the work by Nelander et al.~\cite{Nelander2008}, Molinelli et al. \cite{Molinelli2013}, and Korkut et al.~\cite{Korkut2015} and consider the response of a network to systematic perturbations to the steady state. How the nodes of a regulatory network respond to a perturbation depends on the interactions between nodes. The response to such a perturbation thus encodes information on the underlying regulatory network, and this information can in principle be used to infer the network.

A perturbation is caused by altering the activity of a node in a controlled way. 
In a gene regulatory network, one can disable the production of a particular gene product (a so-called knock-out) or reduce the concentration of a gene product 
(a knock-down)~\cite{hu2007genetic,schubert2018perturbation,kemmeren2014large}. 
This affects the concentrations of other gene products, namely those that are (directly or indirectly) regulated by the perturbed gene. In signalling networks, one can use pharmacological interventions 
to inhibit specific kinases using so-called kinase inhibitors~\cite{dorel2018modelling,Molinelli2013}.
In neural networks, electrical and optoelectrical techniques have been developed to deliver perturbations~\cite{wurtz2015using,chettih2019single}, as well as more drastic and irreversible procedures such as punctate chemical lesions.

Our approach expands previous work to establish a systems biology of perturbations by (i) integrating stochastic fluctuations into the statistical observables used for inference and (ii) constructing an approximate likelihood-based scheme based on these fluctuations. The motivation for incorporating fluctuations into the network inference framework partly comes from the emergence of single-cell experiments, which will provide a rich source of data containing many sources of fluctuations both extrinsic and intrinsic. 

This paper is structured as follows: In section~\ref{modelandsteadystate} we set up a stochastic regulatory model based on a multivariate Langevin equation and use It\^{o}'s lemma to characterize its steady state. In section~\ref{section:Forward problem}, we use the Gaussian mean-field theory~\cite{Mezard2011} developed by M\'ezard and Sakellariou to find self-consistent equations for means and correlations of the activities. These equations solve the forward problem, that is, they determine steady-state observables given the regulatory interactions and other model parameters. In section~\ref{section:Inverse problem}, we then address the inverse problem using both moment matching and a maximum likelihood approach. 
Finally, in section~\ref{syntheticandcelllinesdata} we apply our method to both synthetic data and protein levels measured in a melanoma cell line under different drug treatments~\cite{Molinelli2013}.

\section{Stochastic model of gene regulation}
\label{modelandsteadystate}

Our model of a regulatory network is defined by the coupled stochastic differential equations
\begin{align} \label{eqn:stochastic_differential_equations}
\frac{ \text{d} x_i}{\text{d} t} = a_i \phi \left( h_i \right) - b_i x_i + c_i \xi_i ,
\end{align}
with
\begin{align}
h_i(t) = \sum_{j} \omega_{ij} x_j(t) - \theta_i \, .
\label{eqn:definition_external_field}
\end{align}
$x_i(t)$ denotes the activity of node $i$ at time $t$. In the case of gene regulation, the activity describes the (log-) concentration of a gene product, in the case of a neural network it can describe the firing rate of a neuron. $h_i(t)$ is called a local field, in the context of neural networks it is called the synaptic field.

The first term in \eqref{eqn:stochastic_differential_equations}, $\phi(h)$, is called the transfer function and specifies how the activity of a node is affected by the activities of other nodes. We will discuss specific choices for the transfer function below. The transfer function increases monotonously with the local effective field $h_i$ defined by equation
\eqref{eqn:definition_external_field}. The local effective field acting on node $i$ is a weighted sum of the activities of other nodes $j$. Each weight $\omega_{ij}$ encodes a regulatory interaction between node $j$ and node $i$. A positive weight encodes an activating interaction: a high activity of node $j$ leads to a positive rate of change of the activity of node $i$. A negative weight encodes an inhibitory interaction. Crucially, no symmetry is imposed on these weights. As a result, the steady state of the stochastic dynamics~\eqref{eqn:stochastic_differential_equations} is in general a non-equilibrium steady state.

The local effective field $h_i$ also depends on a threshold $\theta_i$. In the context of neural networks it is called the synaptic threshold. Later, we will consider perturbations of the steady state caused by altering these thresholds to infer the regulatory interactions.

The second term in \eqref{eqn:stochastic_differential_equations} describes an exponential decay of the activity of each node with a decay constant $b_i$. The third term describes stochastic white noise with the standard properties.

Equations \eqref{eqn:stochastic_differential_equations} and \eqref{eqn:definition_external_field} encompass several models in the literature. For a linear choice of the transfer function $\phi(h)$, they define a multivariate Ornstein-Uhlenbeck process~\cite{gardiner2009stochastic}. This linear model has been used to quantify the interactions between network nodes on the basis of steady state correlations~\cite{barzel2009quantifying} and to reconstruct regulatory networks~\cite{gardner2003inferring,bonneau2006inferelator}. With a sigmoid transfer function, they describe a stochastic version of a deterministic model which has been used extensively for gene expression clustering and for the inference of gene regulatory interactions from gene expression data~\cite{mjolsness1999coexpression,weaver1999modeling,bonneau2006inferelator,Nelander2008}.
 With a sigmoid transfer function, they also describe a stochastic variant of the neural Wilson--Cowan model~\cite{Kilpatrick2013} with individually resolved neurons.

\section{The forward problem} \label{section:Forward problem}

The forward problem is to characterize statistical observables, like mean activities $m_i$  and their covariances $\chi_{ij}$ in the steady state, given the model parameters. We discuss two distinct approaches to the forward problem defined by the stochastic model \eqref{eqn:stochastic_differential_equations} -- \eqref{eqn:definition_external_field}. In the first approach, we analyse the stochastic differential equations, in the second approach we apply a Gaussian mean-field theory.

\subsection{Steady-state moments}

The expectation values of the activities and their covariances are denoted
\begin{align}
m_i &= \left< x_i\right> \equiv
\int \text{d}\mathbf{x} \mathcal{P}(\mathbf{x}) x_i \\
\chi_{ij} &= \left< x_i x_j \right> - \left< x_i \right> \left< x_j \right> \equiv
\int \text{d}\mathbf{x} \mathcal{P}(\mathbf{x}) x_i x_j - m_i m_j \, ,
\label{eqn:definition_first_and_second_moment}
\end{align}
where $\mathcal{P}(\mathbf{x})$ is the (generally unknown) multivariate probability distribution over the activities in the steady state. To calculate these observables in the steady state, we rewrite the stochastic differential equation \eqref{eqn:stochastic_differential_equations} as
\begin{align} \label{eqn:ddtx}
\ddt x_i = F_i(\mathbf{x}) +	c_i \xi_i \, ,
\end{align}
where the first term on the right-hand side describes the deterministic contribution to the dynamics
\begin{align} \label{eqn:definition_F}
F_i(\mathbf{x}) = a_i \phi \left( \sum_{ k} \omega_{ik} x_k - \theta_i \right) - b_i x_i \, .
\end{align}
To calculate the steady-state average $\left< x_i \right>$, we average the dynamics \eqref{eqn:ddtx} over the steady-state measure $\mathcal{P}(\mathbf{x})$. Setting $\left< F_k(\mathbf{x}) \right> =0$ in the steady state yields
\begin{equation}
\label{eq:steadystate1}
\left< x_i \right> = \frac{ a_i}{ b_i} \left< \phi \left( h_i \right) \right> \, .
\end{equation}

For the second moments $\left< x_i x_j \right>$, we  start with the total time derivative of the function $f_{ij}( \mathbf{x}) = x_i x_j$. We use the multivariate version of It\^{o}´s lemma~\cite{gardiner2009stochastic} on that function to obtain
\begin{align} \label{eqn:ddtxx}
\begin{split}
\ddt \, ( x_k x_l) & = F_k(\mathbf{x})x_l + F_l(\mathbf{x})x_k \\
& \quad+ \delta_{kl} c_l^2 + c_k x_l  \xi_k + c_l x_k  \xi_l\, .
\end{split}
\end{align}
Again, we perform the average with respect to the steady-state measure on the left- and right-hand sides and set the time derivative to zero yielding
\begin{align}
\begin{split}
0 &= \left< F_k(\mathbf{x})x_l + F_l(\mathbf{x})x_k \right> + \delta_{kl} c_k^2 \\
 \left< x_i x_j \right> &= \frac{ a_i}{ b_i + b_j} \left< \phi \left( h_i \right) x_j \right>
+ \frac{1}{2} \frac{c_i^2}{  b_i + b_j} \delta_{ij} + \left( i \leftrightarrow j \right) \, ,
\end{split}
\label{eqn:steady_state_average}
\end{align}
where we used $\left< x_l  \xi_k \right> = 0 $ for a non-anticipating variable $x_l$. The term $\left( i \leftrightarrow j \right)$ indicates a summand with interchanged indices.

For completeness, we also derive the third moment $\left< x_i x_j x_k \right>$, but will not use it in the following. The calculation
proceeds analogously to the first two moments, with It\^{o}'s lemma applied to the function $f_{ijk}( \mathbf{x}) = x_i x_j x_k$.
Collecting the results, the first three moments of the steady state distribution are given by
\begin{align}
m_i&=\left< x_i \right> = \frac{ a_i}{ b_i} \left< \phi \left( h_i \right) \right> \label{eqn:exact_relations1} \\
C_{ij}&=\left< x_i x_j \right> = \frac{ a_i}{ b_i + b_j} \left< \phi \left( h_i \right) x_j \right> + \frac{1}{2}\frac{c_i^2}{  b_i + b_j} \delta_{ij} + \left( i \leftrightarrow j \right) \label{eqn:exact_relations2}\\
L_{ijk} &= \left< x_{i} x_\text{j} x_{k} \right> = \frac{ a_{i}}{ b_{i} + b_{j} + b_{k}}  \left< \phi \left( h_{i} \right) x_{j} x_{k} \right> \nonumber\\
&+ \frac{c_{i}^2}{ b_{i} + b_{j} + b_{k}} \delta_\text{ij} \left< x_{k} \right> 
+ \left( i \leftrightarrow j \leftrightarrow k \right)  \label{eqn:exact_relations3}\ .
\end{align}

Conceptually, equations \eqref{eqn:exact_relations1} and \eqref{eqn:exact_relations2} are non-equilibrium analogs of Callen's identities~\cite{callen1963note} for the equilibrium state of spin models
(see~\cite{Nguyen2017} for a straightforward derivation). The covariance 
$\chi_{ij} \equiv \left<  x_i x_j \right>-\left<  x_i\right>\left< x_j \right>$
is accordingly given by
\begin{align}
\chi_{ij} &= \frac{ a_i}{ b_i + b_j} \left< \phi \left( h_i \right) x_j \right> + \frac{1}{2}\frac{c_i^2}{  b_i + b_j} \delta_{ij} \nonumber\\
&\quad - \frac{1}{2} \frac{ a_i}{ b_i} \left< \phi \left( h_i \right) \right> \frac{ a_j}{ b_j} \left< \phi \left( h_j \right) \right> + \left( i \leftrightarrow j \right) 
\label{eqn:exact_relations_chi} \ .
\end{align}

\subsection{Gaussian mean-field theory}

We derive a set of self-consistent equations for the steady-state observables $m_i$ and $\chi_{ij}$. To this end, 
we use the Gaussian mean-field theory developed by M\'ezard and Sakellariou in the context of the kinetic Ising model~\cite{Mezard2011}. The key assumption of this theory is that the local effective field defined by equation~\eqref{eqn:definition_external_field} is a Gaussian distributed random variable, with a node-dependent mean and a variance.

The basis for this assumption is the law of large numbers. The local effective fields are asymptotically Gaussian distributed provided the terms in the sum over $j$ in ~\eqref{eqn:definition_external_field} are statistically independent. This means that the effective fields are approximately Gaussian distributed provided that (i) each node $i$ is coupled to many other nodes via finite values of the couplings $\omega_{ij}$ of comparable magnitude, and (ii) the regulatory network encoded by the matrix $\omega_{ij}$ contains no short loops (which would lead to correlations).

To self-consistently determine the mean and variance of the local effective field, we follow M\'ezard and Sakellariou and split each field $h_i$ into a deterministic contribution and a random contribution
\begin{equation} \label{eqn:gaussian_external_field}
h_i( \mathbf{m}) = g_i( \mathbf{m}) + \eta_i \quad
\, ,
\end{equation}
with the deterministic contribution $g_i( \mathbf{m}) = \sum_k \omega_{ik} m_k - \theta_i$.
The random contribution $\eta_i$ follows a multivariate normal distribution  $\mathcal{P}(\{ \eta_i\})$ characterized by zero means and a symmetric covariance matrix $\Delta_{ij}$ to be determined.\\

To obtain self-consistent equations for the mean activity $m_i$ of a given node $i$, we start with the steady-state relation for the mean activity \eqref{eqn:exact_relations1}. Performing the average over $\eta_i$, we obtain the integral equation
\begin{equation}
\label{eqn:self_consistent_equation_m}
m_i  = \frac{1}{\sqrt{2 \pi}} \frac{ a_i}{ b_i} \int \text{d} \eta_i \exp \left( \frac{ {\eta_i}^2}{ 2 \Delta_{ii} } \right) \phi \left( g_i( \mathbf{m}) + \eta_i \right) \, .
\end{equation}
Solving this integral equation requires the variances $\Delta_{ii}$ of the random contribution to the local field $h_i$.
Based on the definition of the local field \eqref{eqn:gaussian_external_field}, the relation between covariance of local fields $\Delta_{ij}$ and covariance of the expression levels $\chi_{ij}$ is given by
\begin{align} 
\Delta_{ij} &= \cov ( \eta_i, \eta_j) = \cov ( h_i, h_j) \nonumber\\
& \stackrel{\eqref{eqn:gaussian_external_field}}{=} \left<  \sum_k \omega_{ik} \left( x_k - m_k\right) \sum_l \omega_{jl} \left( x_l - m_l \right) \right> \nonumber\\
&= \sum_{kl} \omega_{ik} \omega_{jl} \chi_{kl} = \left[ \boldsymbol{\omega \chi \omega}^\intercal \right]_{ij} \label{eqn:covariance_external_field} \, .
\end{align}

To derive a set of self-consistent equations also for the covariances $\chi_{ij}$, we consider the steady state averages $\left< \phi \left( h_i \right) x_j \right>$. 
We invert the interaction matrix
\begin{align} \label{eqn:expression_levels_through_external_fields}
x_i = \sum_k \omega_{ik}^{-1} ( h_k + \theta_k) 
\end{align}
and rewrite the steady-state average
\begin{equation}
	\label{eqn:state_state_average_tanh_h_x}
\left< \phi \left( h_i \right) x_j \right> \stackrel{ \eqref{eqn:expression_levels_through_external_fields}}{=}
\sum_k \omega_{jk}^{-1}  \left< \left( h_k + \theta_k \right) \phi \left( h_i \right) \right> 
\end{equation}
as a sum over averages over local fields, which follow a Gaussian distribution. To calculate the steady-state averages in \eqref{eqn:state_state_average_tanh_h_x}, we define the correlation coefficient for the local fields (Pearson's correlation coefficient)
\begin{align}
\rho_{jk} = \frac{ \cov ( h_j, h_k)}{ \sqrt{ \Delta_{jj} \Delta_{kk}}}  \, .
\end{align}
Following~\cite{Mezard2011} we expand the multivariate normal distribution of the local fields to linear order in the correlation coefficient $\rho_{jk}$. This assumes that the correlation coefficients are small; for fully connected networks they are of order $1/\sqrt{N}$~\cite{Mezard2011}. Within the framework of the Gaussian theory this is a self-consistent assumption as many connections per node also result in a Gaussian local field
\begin{equation}
	\label{eqn:expanded_probability_distribution}
\mathcal{P}( \eta_j, \eta_k) =
 \mathcal{P}( \eta_j) \mathcal{P}( \eta_k) \left( 1 + \frac{\eta_j}{\sqrt{ \Delta_{jj}}} \frac{\eta_k}{ \sqrt{ \Delta_{kk}}} \rho_{jk} \right) + \mathcal{O}(\rho_{jk}^2) \ .
\end{equation}
This gives an approximate expression for the distribution of the local fields, from which we obtain for the averages over effective fields in $\eqref{eqn:state_state_average_tanh_h_x}$
\begin{align} \label{eqn:steady_state_average_tanh_h_h}
\left< h_k \phi \left(  h_j \right) \right> \stackrel{\eqref{eqn:expanded_probability_distribution}}{=} g_k m_j + \lambda_j \cov (h_{k} h_{j}) .
\end{align}
The factor $\lambda_j$ is a measure for sensitivity of the first moments to fluctuations in the activities and is given by the integral equation
\begin{equation} \label{eqn:integral_equation_lambda}
\lambda_j = \frac{1}{\sqrt{2 \pi}} \int \text{d} \eta \exp \left( \frac{ \eta^2}{ 2 \Delta_{ii} } \right)
\phi'\left( g_j + \eta \right)  \, .
\end{equation}

Using the steady-state averages \eqref{eqn:state_state_average_tanh_h_x} and \eqref{eqn:steady_state_average_tanh_h_h} we obtain for the covariances $\chi_{ij}$ of the activities defined in \eqref{eqn:definition_first_and_second_moment}
\begin{align}
\chi_{ij} &\stackrel{\eqref{eqn:definition_first_and_second_moment}}{=} \frac{ a_i}{ b_i + b_j} \left< \phi \left( h_i \right) x_j \right>
+ \left( i \leftrightarrow j \right) + \frac{c_i^2}{  2b_i} \delta_{ij} - \left< x_i \right> \left< x_j \right> \nonumber\\
&\!\!\!\!\!\stackrel{ \eqref{eqn:state_state_average_tanh_h_x} \eqref{eqn:steady_state_average_tanh_h_h}}{=} \frac{ a_i}{ b_i + b_j} \sum_k \omega_{ik}^{-1}
\left( g_k m_j + \lambda_j \cov (h_{k} h_{j}) + \theta_k m_j \right) \nonumber\\
&\quad+ \left( i \leftrightarrow j \right)  + \frac{c_i^2}{2 b_i} \delta_{ij} - m_i m_j \, .
\end{align}
With the definition of $g_i( \mathbf{m})$ in equation \eqref{eqn:gaussian_external_field} and the covariance of the local field in equation \eqref{eqn:covariance_external_field}
we finally obtain a set of self-consistent equations for the covariance matrix of the activities
\begin{align} \label{eqn:self_consistent_equation_chi}
\chi_{ij} &= \frac{a_j}{ b_i + b_j} \left( \chi \omega ^\text{T} \right)_{ij} \lambda_j + \left( i \leftrightarrow j \right) + \frac{c_i^2}{2 b_i} \delta_{ij} \, .
\end{align}

To obtain a solution of the forward problem within Gaussian mean-field theory, we iteratively solve the self-consistent equations for the mean activities \eqref{eqn:self_consistent_equation_m} and their covariances \eqref{eqn:self_consistent_equation_chi} given the model parameters.
In each iteration step, we solve the integrals in \eqref{eqn:self_consistent_equation_m} and  \eqref{eqn:integral_equation_lambda} numerically using adaptive Gauss-Kronrod quadrature \cite{Laurie1997}.
We find that this simple iterative procedure with initial values $\mathbf{m}_\text{init} = \mathbf{0}$ and $\chi_\text{init} = \mathbb{I}$ converges for all model parameters we tried.

\section{The inverse problem}
\label{section:Inverse problem}

We now turn to the inverse problem on the basis of the response of the activities to specific perturbations and ask which set of model parameters reproduces the observed data for all those perturbations. We first consider the standard method of moment matching by least squares\cite{Nelander2008, Molinelli2013, Korkut2015} and then develop an alternative approach based on maximum likelihood. In both approaches, we use the results derived for the forward problem above to derive the mean activities and their correlations as a function of the perturbations and the model parameters.

Each perturbation is modelled as a change in the thresholds in the local effective fields \eqref{eqn:definition_external_field}
\begin{equation}
\theta_i^\mu=\bar{\theta}_i - u_i^{\mu} \ ,
\end{equation}
where $\bar{\theta}_i$ is constant and $u_i^{\mu}$ is different for each perturbation.
The different perturbations are labelled $\mu = 1,2,\ldots,M$ and one or several measurements of the activities $\{x_i\}$ are taken per perturbation. A large positive value of $u_i^{\mu}$ means that gene $i$ is artificially upregulated in a given perturbation $\mu$, a large negative value means that that gene is downregulated.

The model parameters consist of the regulatory interactions $\omega_{ij}$, the thresholds $\bar{\theta}_i$, as well as the prefactors of the transfer function $\{a_i\}$, the decay constants $\{b_i\}$, and the amplitudes of the stochastic noise $\{c_i\}$ (see \eqref{eqn:stochastic_differential_equations}). However, rescaling all parameters $a_i$ and
$b_i$ by the same factor, and the amplitudes of the noise $c_i$ by the square root of that factor leaves the observables (means and covariances) unchanged. Hence, the model para\-meters can only be inferred up to a global constant, which acts like a gauge parameter for the inference. Changing this parameter corresponds to rescaling the unit of time in the multivariate Langevin equation~\eqref{eqn:stochastic_differential_equations}.Since the steady state observables are time-independent, this scale factor cannot be recovered from the steady state. 

\subsection{Moment matching by least squares} \label{subsection: least-squares methods}

We start with a least-squares fit based on the average activities in the steady state given by \eqref{eqn:exact_relations1}. A quadratic cost function penalizing the difference between the observed average activities and those given by is~\eqref{eqn:exact_relations1}
\begin{align}
	\label{eq:ms1o}
	\lambda_\text{ms1o}(\boldsymbol{\omega},\boldsymbol{\theta},\mathbf{a},\mathbf{b})=
 \sum_{ \mu i} \left[ \vphantom{\frac{ a_i}{ b_i}} \left< x_i^\mu \right>
- \frac{ a_i}{ b_i} \left< \phi \left( h_i^\mu( \mathbf{x}^\mu, \boldsymbol{\omega}, \theta^\mu_i) \right) \right>  \right]^2  \ . 
\end{align}
$x^\mu_i$ denotes the activity of node $i$ under perturbation $\mu$, and we assume for now that there are multiple experimental measurements per perturbation (replicates), over which expectation values can be computed.
The corresponding estimate of the model parameters is obtained by minimizing the quadratic cost function with respect to the parameters~\cite{Nelander2008, Molinelli2013, Korkut2015}.

In the next step, we incorporate also the information contained in the covariances of the activities into the fit. To this end, we penalize differences between the observed
mean activities as well as their covariances with the corresponding quantities in the steady state, leading to a quadratic cost function given by
\eqref{eqn:exact_relations1} and \eqref{eqn:exact_relations2}
\begin{align}
		\label{eq:ms2o}
\begin{split}
&\lambda_\text{ms2o}(\boldsymbol{\omega},\boldsymbol{\theta},\mathbf{a},\mathbf{b},\mathbf{c}) = \sum_{ \mu i} \left[ \vphantom{\frac{ a_i}{ b_i}} \left< x_i^\mu \right>-
 \frac{ a_i}{ b_i} \left< \phi \left( h_i^\mu( \mathbf{x}^\mu, \boldsymbol{\omega}, \theta^\mu_i) \right) \right> \right]^2 \\
& \quad +  \frac{1}{N} \sum_{ \mu {ij}} \left[ \left< x_i^\mu \, x_j^\mu \right> - \left< x_i^\mu \right> \left< x_j^\mu \right> - \chi_{ij} \left( \mathbf{x}^\mu, \boldsymbol{\omega},\boldsymbol{\theta}^\mu,\mathbf{a},\mathbf{b},\mathbf{c}\right) \right]^2  .
\end{split}
\end{align}
In the second line, we used the steady-state result \eqref{eqn:exact_relations_chi} for the covariance 
$\chi$. Due to the scaling of the number of mean activities and the number of covariances with the number of nodes $N$ of the regulatory network (linear and quadratic, respectively), we used a relative factor of $1/N$ weighting these two terms in \eqref{eq:ms2o}.

We can also use the Gaussian mean-field theory developed in section \ref{section:Forward problem} for the mean activities $m_i( \boldsymbol{\omega})$ and their covariances $\chi_{ij}$ for moment matching.
The resulting quadratic cost function for matching the first two moments is
\begin{align}
	\label{eq:msGt}
		\lambda&_\text{msGt}(\boldsymbol{\omega},\boldsymbol{\theta},\mathbf{a},\mathbf{b},\mathbf{c}) =  \sum_{ \mu i} \left[ \left< x_i^\mu \right> - m_i^\mu(\boldsymbol{\omega},\boldsymbol{\theta},\mathbf{a},\mathbf{b},\mathbf{c})  \right]^2 \\
		&+ \frac{1}{N} \sum_{ \mu {ij}} \left[ \left< x_i^\mu \, x_j^\mu \right> - \left< x_i^\mu \right> \left< x_j^\mu \right> -  \chi_{ij}(\boldsymbol{\omega},\boldsymbol{\theta}^{\mu},\mathbf{a},\mathbf{b},\mathbf{c}) \right]^2  \, . \nonumber
\end{align}

\subsection{Maximum likelihood} \label{subsection: Maximum likelihood method}

An alternative to moment matching is parameter inference by maximum likelihood. 
This has a number of attractive properties compared to moment matching. In the limit of many sets of input data, the maximum likelihood estimator converges in probability to the underlying model parameter (the consistency property). Furthermore, for large sample sizes, there is no consistent estimator which results in a smaller mean-squared error. Most importantly, maximum likelihood does not rely on moments
as input, so it can be used even when there are only single measurements taken per perturbation.  

However, the exact likelihood function is not known in our case, since we do not know the distribution $\mathcal{P}(\mathbf{x})$ of activities in the non-equilibrium steady state. Instead, we use the simplest approximation possible for the steady state; a multivariate Gaussian distribution whose means and variances are specified by the results of the forward problem solved in section~\ref{section:Forward problem}.
This approximate likelihood of the model parameters given the activities $\mathbf{x}^\mu$ is denoted by $\mathcal{P}_g( \mathbf{x}^\mu ,\boldsymbol{\omega}, \mathbf{\theta},\mathbf{a}, \mathbf{b}, \mathbf{c})$.
The resulting log-likelihood function
\begin{align} \label{eqn:mlGt}
\begin{split}\ln &\mathcal{P}_\text{g}(\mathbf{x}^\mu, \boldsymbol{\omega},\boldsymbol{\theta},\mathbf{a},\mathbf{b},\mathbf{c}) \\
	&= - \frac{1}{2} \sum_{ \mu} \left( \mathbf{x}^\mu - \mathbf{m} \right)^\text{T}  \chi^{-1} \left( \mathbf{x}^\mu - \mathbf{m} \right) - \vphantom{ \sum_{ \mu k}} \frac{1}{2} \ln \det \chi   \, ,
\end{split}
\end{align}
depends on the solution of the self-consistent equations \eqref{eqn:self_consistent_equation_m} for $\mathbf{m}=\mathbf{m}( \boldsymbol{\omega},\boldsymbol{\theta}^\mu,\mathbf{a},\mathbf{b},\mathbf{c})$ and  \eqref{eqn:self_consistent_equation_chi} for $\boldsymbol{\chi}=\boldsymbol{\chi} \left(  \boldsymbol{\omega},\boldsymbol{\theta}^\mu, \mathbf{a}, \mathbf{b}, \mathbf{c} \right)$. The index $\mu$ in this context refers to the different experimental samples.

Like the quadratic cost function \eqref{eq:ms1o}, the approximate likelihood function \eqref{eqn:mlGt} also quadratically penalizes the difference between the mean activities in the data and those expected under a forward model. However, in the maximum likelihood function \eqref{eqn:mlGt}, this penalty depends on the variance of the activities, and is lower for nodes with high variance, whose activities are in a sense less informative as a result.
Since the approximate likelihood function \eqref{eqn:mlGt} requires the means and covariances as a function of the model parameters, it is necessary to use the Gaussian mean-field theory (or an equivalent approach) here, rather than equations ~\eqref{eqn:exact_relations1} and~\eqref{eqn:exact_relations2}. 

\subsection{Comparison of least-squares fitting and maximum likelihood} \label{subsection: Comparison of least-squares and maximum likelihood method}

\subsubsection{Inference from artificial perturbation data}

We first use simulated data to compare the results of the least-squares methods and the maximum-likelihood approach. To produce simulated data we chose the transfer function $\phi(h)=\tanh(h)$
used in Nelander et al.~\cite{ Nelander2008} as a model for the (log-)expression levels in a gene regulatory network.
To generate a gene interaction matrix $\boldsymbol{\omega}$, we set $ \omega_{ij} = \beta \zeta_{ij} \forall i \neq j $, where $\zeta_{ij}$ is a random variable drawn independently from a normal distribution with mean equal to zero and variance equal to the inverse of total numbers of genes, $1/N$. Self-regulation is not considered, $\omega_{ii} = 0 \forall i \neq j $. The inter-gene coupling strength $\beta$ (set to $1$ here) is a measure of the regulation strength between genes. We use perturbations of single nodes where for every perturbation one gene is perturbed with $u^\mu_i = -2$. The other model parameter were set to $\bar{\theta}_i = 0$, $a_i = 1$, $b_i = 1$, and $c_i = 1$.

To quantify the quality of the network inference we measure the reconstruction error
\begin{equation}
r = \sqrt{ \frac{ \sum_{{ij}} \left( \omega_{ij}^\text{gen} - \omega_{ij}^\text{rec} \right) ^ 2}{ \sum_{ {ij}} \left( \omega_{ij}^\text{gen} \right)^2 }} \label{eq:reconstructionerror} \ ,
\end{equation}
which gives the relative mean-square difference between the inferred interactions and the interactions which were used to generate the data.

\begin{figure}
	(A) \includegraphics{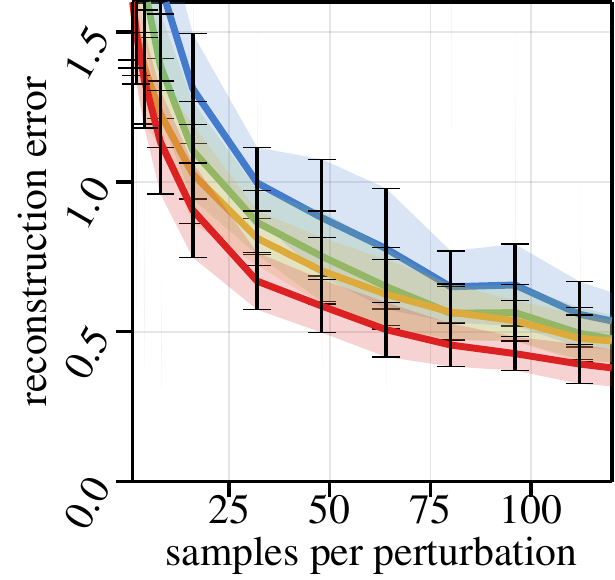} \\ \vspace{0.5cm}
	(B)
	\includegraphics{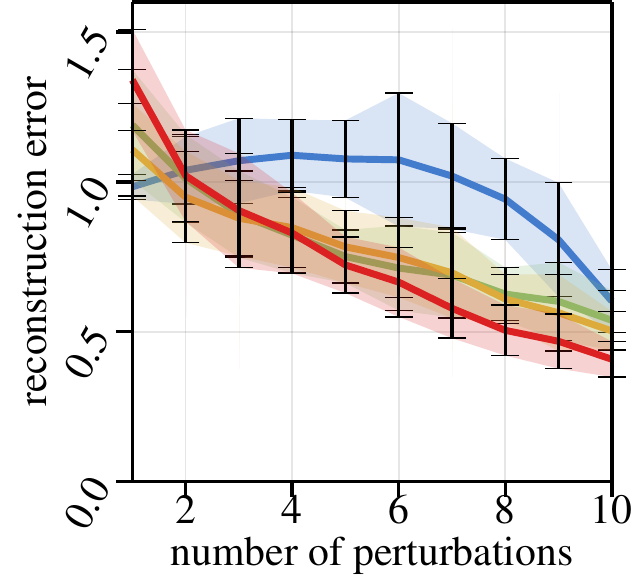}
	\centering
	\caption{ (A) The reconstruction error \eqref{eq:reconstructionerror} of the regulatory interactions is plotted against the number samples per perturbation for a fixed number of $10$ single-drug perturbations. (B) The reconstruction error is plotted against the number of single-drug perturbations and the inference is based on $100$ samples per perturbation. The system size is $N=10$ in both cases, and the model parameters are as described in the main text. The least-squares fit of the first moments~\eqref{eq:ms1o} (blue), the second moments~\eqref{eq:ms2o} (green), and second moment fit based on the Gaussian mean-field theory~\eqref{eq:msGt} (yellow) are shown we well as the approach based on maximum likelihood (red), which outperforms the other approaches.}
	\label{plots:reconstruction_error_c=1.0}
\end{figure}

Figure \ref{plots:reconstruction_error_c=1.0} shows that, as expected, the inference of regulatory interactions is improved over matching the first moments using~\eqref{eq:ms1o} by matching the second moments using~\eqref{eq:ms2o} as well. Using the Gaussian mean-field theory to compute the moments rather than using the expressions~\eqref{eqn:exact_relations1} and~\eqref{eqn:exact_relations2} makes no significant difference to the inference. However, the approximate maximum likelihood~\eqref{eqn:mlGt} clearly outperforms the least-squares method. 

However, the maximum likelihood turns out to be more costly computationally. Both approaches are based on iteratively solving self-consistent equations, so there is no easy way to estimate differences in running times of the two approaches. 
 For a small network consisting of $10$ nodes reconstruction by maximizing the likelihood takes about $60$ seconds on a standard processor, compared to about $5$ seconds for moment matching. We used the local optimization routine LN\_BOBYQA from the NLopt package with starting points $\omega_{ij} = 0$. We found that global optimization routines, like GN\_DIRECT\_L, GN\_CRS2\_LM, GN\_ESCH from the NLopt package, did not yield better results. We expect that advanced algorithms, like message passing~\cite{Molinelli2013,Korkut2015} can speed up the reconstruction process, but are outside the scope of this paper. 
 
 Also, we did not repeat the optimization process many times over to determine interactions which appear in many near-optimal choices of the model parameters~\cite{Molinelli2013}, as our focus is on the construction and comparison of different cost functions. The restriction to a few high-confidence interactions can also be achieved by a suitable regularizing function, which is also outside the scope of this paper. 

\subsubsection{Inference from cell line perturbation data}
\label{syntheticandcelllinesdata}

SK-MEL-133 is a human cancer cell line taken from a melanoma patient and grown in culture.
Molinelli et al.~\cite{Molinelli2013} report perturbation experiments on this cell line, 
which we use for network inference. The perturbation experiments focus on 
the PI3K/AKT and MAPK pathways, which play a key role in several different types of cancers. For $16$ proteins of the PI3K/AKT and MAPK pathways, protein levels (total or phospho-levels) are quantified using reverse phase protein arrays; these $16$ proteins form the nodes of the network to be inferred. Each level is specified as a log-2 ratio of the perturbed and the unperturbed protein 
level, see~\cite{Molinelli2013} for details. We note that the reverse phase protein arrays take information from a bulk of cells, and correspondingly fluctuations and co-variance observed in that data do not come from the stochastic expression of genes in single cells (see discussion). 

The perturbations are implemented using $8$ different compounds (inhibitors), which inhibit the signalling activity of particular kinases that are part of the PI3K/AKT and MAPK pathways.
$8$ inhibitors are used singly and in pairwise combinations to perturb the pathways. Altogether, $44$ different perturbations are used. Each inhibitor is given at the concentration where it reduced the growth of the cell line by $40\%$, the so called $\text{IC}_{40}$ value. For several of the inhibitors used, the targets lie upstream of one of the $16$ proteins, so the protein is not targeted directly, but the downstream effector of each compound is known. We set those known interactions between the $8$ inhibitors and the $16$ proteins equal to one (figure 5A in \cite{Molinelli2013}), and set all other interactions between inhibitors and proteins equal to zero. 

We used the $44$ perturbations to generate $8$ training sets and $8$ matching testing sets. Each training set contains all the single-compound and multi-compound combinations which do not contain one specific compound $d$. The matching testing set contains only the single-compound perturbation using compound $d$. This is done to ensure that no information on compound $d$ is included during training, beyond what can be inferred about the network of interactions between nodes. Thus, a prediction about the activities of the nodes under compound $d$ must come from the inferred information about the network~\cite{Molinelli2013,Korkut2015}.

Based on each training set, we inferred the regulatory interactions $\boldsymbol{\omega}$, as well as the parameters $\mathbf{a}$, $\mathbf{b}$, and $\mathbf{c}$. 
A single parameter, $a_1$ with node $1$ corresponding to EIF4EBP1, was set to $1$ throughout in order to fix the single global gauge parameter described in~\ref{section:Inverse problem}. Starting point was $\omega_{ij} = 0$, $a_i=1$, $b_i = 1$, and $c_{i} = 0.1$.
We used the local optimization routine LN\_BOBYQA from the NLopt with starting point $\omega_{ij} = 0$, $a_i=1$, $b_i = 1$, and $c_{i} = 0.1$. Again we found that global optimization routines, like GN\_DIRECT\_L, GN\_CRS2\_LM, GN\_ESCH from the NLopt package, did not yield better results.

For each training set, we then predict the node activities under perturbations in the matching testing set. For the prediction of the activities, we numerically simulated the dynamic model \eqref{eqn:stochastic_differential_equations} for its steady state at the inferred model parameters. This was done to avoid using an approximation (like the Gaussian approximation) as part of the testing.

\begin{figure*}
	(A) \includegraphics{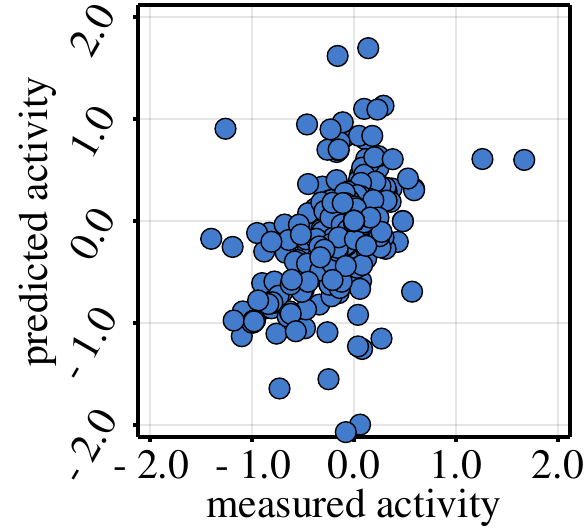} \hspace{0.5cm}
	(B) \includegraphics{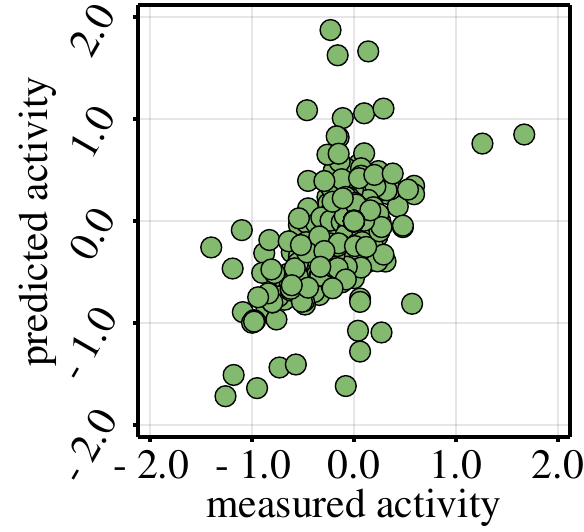}\\ \vspace{0.5cm}
	(C) \includegraphics{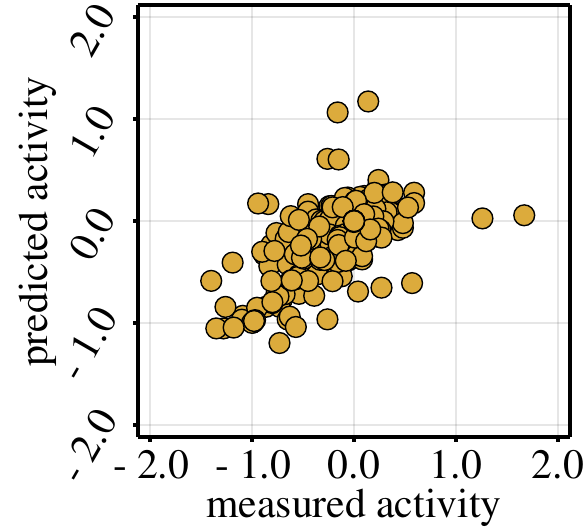} \hspace{0.5cm}
	(D) \includegraphics{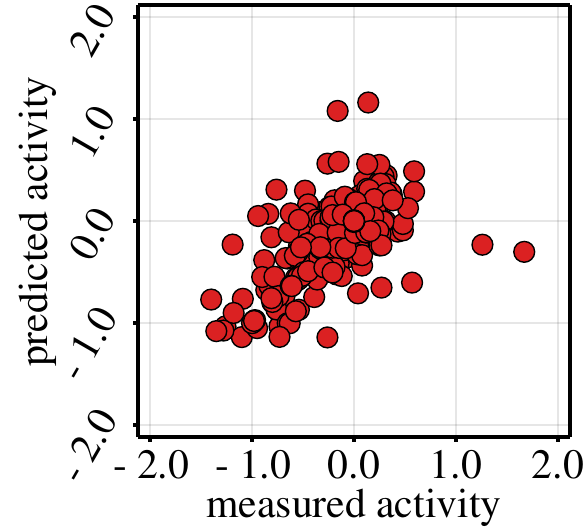}
	\centering
	\caption{Scatter plots comparing the experimentally measured and predicted activities within the SK-MEL-133 cell line experiment (see main text) across different genes and pairs of training and testing set. (A) Prediction based on the least-squares fit in first order \eqref{eq:ms1o}. (B) Prediction based on the least-squares fit in second order \eqref{eq:ms2o}. (C) Prediction based on the least-squares fit within the Gaussian theory \eqref{eq:msGt}. (D) Prediction based on the maximum-likelihood approach \eqref{eqn:mlGt}. The Pearson correlations between measured and predicted activities for A-D are $0.55,0.60,0.69$, and $0.69$, respectively.}
	\label{figure:cellline}
\end{figure*}

Figure \ref{figure:cellline} shows how well the parameters inferred from the training sets model the protein levels under perturbations from the matching testing sets. 
These scatter plots show the measured protein levels (on the $x$-axes) and the predicted expression levels (see above) on the $y$-axes, across different nodes and training/testing sets. The four plots show the predictions based on networks reconstructed using (A) the least-square first order fit, (B) the least-square second order fit, (C) the Gaussian theory, and (D) the maximum-likelihood approach. 
 We find a Pearson correlation of $d_\text{ms1o} = 0.55$ for the least-square first order, $d_\text{ms2o} = 0.60 $ for the least-square second order, $d_\text{msGo} =0.69 $ for the Gaussian theory, and $d_{ml} =0.69$ for the maximum-likelihood method. Including the information on the covariance of activities hence improves the reconstruction, as does the use of the Gaussian theory and the approximate likelihood.

\section{Conclusion}

We have used a stochastic model of regulatory networks for the inference of the regulatory interactions from steady-state perturbation data. To this end, we have extended the moment-matching approach of Nelander et al. and others~\cite{Nelander2008} to fluctuations in the steady state. Based on the stochastic model, we constructed a simple likelihood function for network inference. We found that including information on covariances improved the reconstruction quality both when tested both on artificial and experimental data. The likelihood-based approach also allows using perturbation data when not enough replicates have been measured experimentally to accurately compute expectation values for each perturbation. 

However, the likelihood-based approach comes at a significant computational cost. In our current implementation, the likelihood is maximized via gradient descent, where each step requires solving self-consistent equations for the means and variances of the network's activities. Alternative implementations, for instance message passing, can potentially  increase the speed of execution significantly~\cite{Molinelli2013}. 

Like previous studies, we have used the network's response to perturbations for network inference. The motivation for using perturbations is that 
pairwise correlations (a symmetric matrix) on their own are insufficient for the inference of the network interactions (an asymmetric matrix), and perturbations are one way to enlarge the scope of information extracted from the steady state. However, it is not the only approach: For instance, correlations at different time intervals $\langle x_i(t) x_j(t+\tau)\rangle$ can be calculated and compared with experimental data at different time points without the need for perturbations. In that context, the random fluctuations due to low copy numbers of the molecular machinery of transcription act as a series of perturbations to which the system responds. 
Recently, Gupta et al.~\cite{gupta2022inferring} have studied time-shifted correlations of gene expression levels in single-cell data for network reconstruction. Gupta et al. conclude that, with current technology, the inference of entire regulatory networks on the basis of fluctuations at the single cell-level is not yet feasible. This may soon change, and we expect that replacing  aggregate quantities like correlations with a likelihood-based approach will improve the resulting network inference.

\begin{acknowledgments}

This work was supported by DFG grant CRC1310. 

\end{acknowledgments}


%

\end{document}